\documentclass[doublecol]{epl2}
\usepackage{amssymb}
\usepackage{amsbsy}
\usepackage{amsmath}

\title{Equivalence of Transport Coefficients in Bath-Induced and Dynamical Scenarios}

\shorttitle{Equivalence of Transport Coefficients in Bath-Induced and Dynamical Scenarios}

\author{Robin Steinigeweg\inst{1} \and Marcel Ogiewa\inst{1} \and Jochen Gemmer\inst{1}}

\shortauthor{R. Steinigeweg \etal}

\institute{\inst{1}
Fachbereich Physik,
Universit\"at Osnabr\"uck,
Barbarastrasse 7,
D-49069 Osnabr\"uck, Germany
}


\abstract{
We investigate the transport of a single excitation through a
chain of weakly coupled subunits. At both ends the chain is
exposed to baths which are incorporated by means of a master
equation in Lindblad form. This master equation is solved by the
use of stochastic unraveling in order to obtain excitation profile
and current in the steady state. Completely diffusive transport is
found for a range of model parameters, whereas signatures of
ballistic behavior are observed outside this range. In the
diffusive regime the conductivity is rather independent from the
strength of the bath coupling and quantitatively agrees with the
diffusion coefficient which has been derived from an investigation
of the same model without baths. Also the ballistic behavior in
the non-diffusive regime is in accord with results from this
alternative approach.
}

\begin{document}

\maketitle

There essentially exist two direct approaches to the investigation
of gradient-driven transport phenomena such as, e.g., heat conduction:
({\bf i.})~A closed scenario is considered, where transport is driven
by an internal gradient, i.e., transport is somehow analyzed for the
relaxation of a spatially non-uniform energy distribution (This relaxation
is in accord with a diffusion equation in the case of normal transport.)
\cite{gemmer2006, steinigeweg2007-1, bonetto2000, lepri2003, garrido2001};
({\bf ii.})~An open scenario is considered, where transport is induced by
an external gradient, i.e., baths with different temperatures are locally
coupled to both ends of a system such that a stationary non-equilibrium
state results (with a finite current and a spatially linear energy profile
in the normal transport case) \cite{bonetto2000, lepri2003, garrido2001,
saito2003, michel2008, wichterich2007, benenti2008, mejiamonasterio2005,
prosen2008, benenti2009}.

Furthermore, there are two indirect approaches to gradient-driven
transport: ({\bf iii.})~The Green-Kubo approach \cite{kubo1991} which
essentially gives a diffusion coefficient as an integral over the current
auto-correlation function corresponding to the spatially uniform equilibrium
state; ({\bf iv.})~The Einstein-Herzfeld approach \cite{herzfeld1958} which
relates the diffusion coefficient to the mean square displacement in the
equilibrium state. (For a comprehensive fundamental review of
({\bf iii.})~and ({\bf iv.})~in the context gradient-driven transport see
Ref.~\cite{zwanzig1965} and references therein.)

However, even though all those approaches are commonly expected to
yield equivalent results on transport, direct, say, qualitative
comparisons for concrete quantum models appear to be rare in the
literature \cite{garrido2001, saito2003, michel2008, prosen2008}.
Quantitative ones are even more rare. For the model addressed
below the equivalence of ({\bf i.})~and ({\bf iii.})~has been
investigated and partially confirmed (in the diffusive regime) in
Ref.~\cite{gemmer2006}. Thus, in this Letter we focus exclusively
on the equivalence of ({\bf i.})~and ({\bf ii.}), leaving the
comparison of, say, ({\bf i.})~and ({\bf iv.})~as a relevant
subject of future work in that direction.

The Letter at hand provides such a comparison for a quantum model
which describes the transport of a single excitation through a
finite chain of weakly interacting subunits. This chain has
already been treated as a closed system and is known to exhibit
both purely diffusive and completely ballistic dynamics, depending
on the model parameters \cite{gemmer2006, steinigeweg2007-1}. At
both ends the chain is now exposed to baths which are incorporated
by means of a master equation in Lindblad form \cite{lindblad1976,
breuer2007-1} such that the efficient numerical method of
stochastic unraveling is applicable \cite{dalibard1992,
molmer1996}. For this open system we also find the above regimes
of diffusive and ballistic transport.

In the normal transport case we particularly show the following:
({\bf i.})~The steady state of the open system has a constant
current and a linear excitation profile; ({\bf ii.})~The resulting
conductivity is almost independent from the strength of the bath
coupling; ({\bf iii.})~This conductivity coincides with the
theoretical diffusion coefficient of the closed system, too. The
latter two points deserve closer attention, since it generally is
difficult to ensure that the extracted conductivity is a pure bulk
property, especially for a finite system \cite{bonetto2000,
lepri2003, michel2008}.

In the ballistic transport case we finally demonstrate the
sensitive dependence of the conductivity on the bath coupling
strength. In particular we illustrate that huge, say, infinite
conductivities can be observed solely in the limit of weak bath
couplings, e.g., where the resistance due to the bath contact is
large.

\begin{figure}[htb]
\centering
\includegraphics[width=0.95\linewidth]{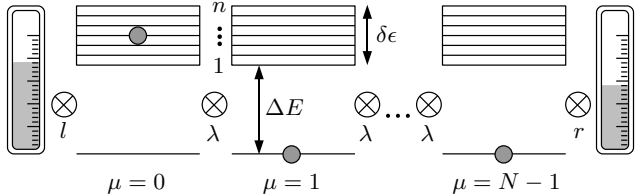}
\caption{Sketch of the considered open system: a chain with a bath
at both ends. The chain consists of $N$ identical subunits which
feature a non-degenerate ground state, a wide energy gap $\Delta
E$, and a narrow energy band $\delta \epsilon$ with $n$
equidistant levels. A state from the investigated one-excitation
subspace is indicated (circles).} \label{model}
\end{figure}

Concretely, we investigate an open quantum system according to
Fig.~\ref{model}: It is an one-dimensional structure which is
connected to baths at both ends. Before the realization of these
baths is discussed below, we specify the chain itself, i.e., we
introduce the closed quantum system at first. It consists of $N$
identical subunits which feature a non-degenerate ground state
$\epsilon_0 = 0$, a wide energy gap $\Delta E$, and a narrow
energy band $\delta \epsilon$ with $n$ equidistant levels
$\epsilon_i = \Delta E + i \, \delta \epsilon / (n-1)$.

In the following the consideration will be focused on the
invariant zero- and one-excitation subspace which is spanned by
the basis $\{ | 0 \rangle, | \mu, i \rangle \}$. In the single
state $| 0 \rangle$ all subunits are in their ground state. In the
$N \, n$ states $| \mu, i \rangle$ only the $\mu$th subunit is
excited to the $i$th level of its band, while all other subunits
are still in their ground state. By the use of this notation the
local Hamiltonian of the $\mu$th subunit may be written as
$\hat{h}_\mu = \sum_{i=1}^n \epsilon_i \, |\mu, i\rangle \langle
\mu ,i |$. The next-neighbor interaction between two adjacent
subunits $\mu$ and $\mu +1$ is supposed to be
\begin{equation}
\hat{v}_\mu = \lambda \sum_{i, j = 1}^n c_{i,j} \, | \mu, i
\rangle \langle \mu + 1, j | \, + \, \text{H.c.}
\end{equation}
with the overall coupling strength $\lambda$. The
$\mu$-independent coefficients $c_{i,j}$ are complex,
independent, and random numbers: their real and imaginary parts
are both chosen corresponding to a Gaussian distribution with mean
$0$ and variance $1/2$. Note that only a \emph{single} realization
of the $c_{i,j}$ (and not some ensemble average over different
realizations) is considered throughout this Letter. The total
Hamiltonian may be written as $\hat{H} = \hat{H}_0 + \hat{V}$,
where $\hat{H}_0$ is the sum of the local Hamiltonians
$\hat{h}_\mu$ and $\hat{V}$ is the sum of the next-neighbor
interactions $\hat{v}_\mu$, respectively.

Of particular interest are the local probabilities $p_\mu(t)$ for
finding an excitation of the $\mu$th subunit somewhere to its
band. These quantities are conveniently expressed as the
expectation values $p_\mu(t) = \text{Tr} \{ \rho(t) \, \hat{p}_\mu
\}$ of respective operators \mbox{$\hat{p}_\mu = \sum_{i = 1}^n
|\mu, i \rangle \langle \mu, i|$}, where $\rho(t)$ denotes the
density matrix of the whole chain. The associated local currents
$j_\mu(t)$ may be defined as the expectation values of the
operators $\hat{j}_\mu = \imath \, [ (\hat{p}_\mu -
\hat{p}_{\mu+1})/2, \hat{v}_\mu ]$, cf.~Ref.~\cite{gemmer2006}.

Even though this system is not meant to represent a concrete
physical situation, it may be illustrated as a simplified model
for a chain of, say, coupled atoms, molecules, or quantum dots. In
this case the hopping of the excitation from one subunit to
another corresponds to energy transport, especially if $\lambda
\ll \Delta E$. It may also be viewed as a model for
non-interacting particles on a lattice with many orbitals per
site. The hopping of the excitation corresponds to transport of
particles in this case. A more detailed discussion of physical
realizations can be found in Refs.~\cite{gemmer2006,
steinigeweg2007-1}.

However, for the closed system it has reliably been shown that the
dynamical behavior of the $p_\mu(t)$ is well described by a
diffusion equation of the form $\dot{p}_\mu(t) = D \, [
p_{\mu-1}(t) - 2 \, p_\mu(t) + p_{\mu+1}(t) ]$, if only the two
conditions
\begin{equation}
\frac{8 \pi \, n \, \lambda^2}{\delta \epsilon^2} \ll 1 \, , \;
\left ( \frac{4 \pi^2 \, n \, \lambda}{N \, \delta \epsilon}
\right )^2 \gg 1 \label{conditions}
\end{equation}
are fulfilled, see Ref.~\cite{steinigeweg2007-1}. The pertinent
diffusion constant is given by $D = 2 \pi \, n \, \lambda^2 /
\delta \epsilon$. Thus, the relaxation time of the excitation
profile's Fourier component with wave length $s$, i.e., the
characteristic time for transport on this length scale, is $\tau_R
= 1/ [ \, 2 \, D (1 - \cos 2 \pi / s) \, ]$.

The first condition (\ref{conditions}) guarantees that the
dynamics on the shortest length scale (s = 1) is much slower than
a generic correlation time as generated by the local parts of the
Hamiltonian: $\tau_C \sim 1 / \delta \epsilon$.

The second condition (\ref{conditions}) ensures that the dynamics
on the longest length scale ($s = N$) is much faster than the
period of any correlation function as generated by the local parts
of the Hamiltonian: $T = 2 \pi \, n / \delta \epsilon$. When this
criterion is broken, e.g., by $\lambda \rightarrow 0$ or $N
\rightarrow \infty$, a transition towards ballistic transport is
caused in the large $s$-limit. In that case the dynamics is still
found to be governed by the above diffusion-type equation but now
with a time-dependent rate, i.e., $D \rightarrow 2 \, D \, t/T$,
cf.~Ref.~\cite{steinigeweg2007-1}.

Our aim is to compare these findings with the results which are
obtained in the following treatment of the same model as an open
system which is exposed to baths. In order to incorporate these
baths we now postulate that a quantum master equation (QME) in
Lindblad form holds, namely,
\begin{equation}
\dot{\rho}(t) = {\cal L} \, \rho(t) = \imath [ \, \rho(t), \hat{H}
\, ] + {\cal D} \, \rho(t) \, . \label{lindblad1}
\end{equation}
The first part on the r.h.s.~of this equation describes the
coherent time evolution of $\rho(t)$ w.r.t.~$\hat{H}$. The second
part is an incoherent damping term and given by
\begin{equation}
{\cal D} \, \rho(t) = \sum_i a_i \left ( \hat{A}_i^{} \, \rho(t)
\, \hat{A}_i^\dagger - \frac{1}{2} [ \, \rho(t), \hat{A}_i^\dagger
\hat{A}_i^{} \, ]_{_+} \right) \label{lindblad2}
\end{equation}
with non-negative rates $a_i$, Lindblad operators $\hat{A}_i$, and
the anti-commutator $[\ldots,\ldots]_{_+}$.
Eqs.~(\ref{lindblad1}), (\ref{lindblad2}) are the most general
form of a linear and time-local QME which defines a trace- and
hermicity-perserving, completely positive dynamical map. This
particularly means that any density is mapped to a density matrix
\cite{breuer2007-1, lindblad1976}.

It is well-known that a strict derivation of a physically
reasonable Lindblad form from a microscopic bath model is somewhat
subtle for conduction scenarios \cite{wichterich2007}. However,
since numerics eventually turns out to be rather involved in that
case, here we primarily choose the computationally least costly
bath implementation (within the established framework of open
quantum systems) which yields a physically acceptable result.
Lindblad forms are computationally preferable, since those allow
for the efficient numerical method of stochastic unraveling
\cite{dalibard1992, molmer1996}. But for our below choice of
Lindblad operators the underlying microscopic formulation can in
principle be found by following the ideas in
Ref.~\cite{wichterich2007}.

Thus, we choose only two left-bath operators
\begin{equation}
\hat{A}_0 \equiv \hat{L}^+ = \frac{1}{\sqrt{n}} \sum_{i = 1}^n |
0, i \rangle \langle 0 | \, , \; \hat{A}_1 \equiv \hat{L}^- =
(\hat{L}^+)^\dagger \label{operators1}
\end{equation}
with rates $a_0 \equiv l^+$, $a_1 \equiv l^-$ but $2 \, n$
right-bath operators
\begin{equation}
\hat{A}_{2i} \equiv \hat{R}^+_i = | N-1, i \rangle \langle 0 | \,
, \; \hat{A}_{2i+1} \equiv \hat{R}^-_i = (\hat{R}^+_i)^\dagger \,
, \label{operators2}
\end{equation}
$i = 1, \ldots, n$ with rates $a_{2i} \equiv r^+$, $a_{2i+1}
\equiv r^-$. $\hat{L}^+$ and $\hat{L}^-$ describe transitions
between the ground state $| 0 \rangle$ and a state where the
excitation is equally distributed over all band levels of the
leftmost subunit. $\hat{R}^+_i$ and $\hat{R}^-_i$ are responsible
for transitions between the ground state and the band state $|
N-1, i \rangle$ of the rightmost subunit. (Note that this contact
modelling goes beyond the secular approximation and is not in
principle conflict with some microscopic picture
\cite{wichterich2007}). We further choose $r^+ = 0$, i.e., we
suppose that the right bath is at zero temperature. The
temperature of the left bath is finite, cf.~Fig.~\ref{model}.

The method of stochastic unraveling relies on the fact that any
Lindblad QME for a density matrix $\rho(t)$ can equivalently be
formulated in terms of a stochastic Schr\"odinger equation (SSE)
for a wave function $| \varphi(t) \rangle$, see
Refs.~\cite{dalibard1992, molmer1996}. This fact advantageously
allows to deal with the same problem in the lower dimensional
Hilbert space. In order to simulate the process which is defined
by the SSE we apply the procedure in, e.g.,
Ref.~\cite{michel2008}: It consists of deterministic evolutions
(w.r.t.~an effective Hamiltonian $\hat{H}_{\text{eff}} = \hat{H} -
\imath/2 \, \sum_i a_i \, \hat{A}_i^\dagger \, \hat{A}_i^{} \,$)
and stochastic jumps (corresponding to one of the Lindblad
operators $\hat{A}_i$). Each application of this procedure leads
to a single realization of a so-called ``trajectory'' $|
\varphi(t) \rangle$. The expectation value of some observable
$\hat{M}$ in the steady state $\rho$ can be evaluated from such a
trajectory by the time average $\text{Tr} \{ \rho \, \hat{M} \}
\approx 1/t \int_0^t \text{d}t_1 \langle \varphi(t_1) \, | \hat{M}
| \, \varphi(t_1) \rangle$ for a sufficiently large time interval
$t$ \cite{molmer1996}. We are primarily interested in $\hat{M} =
\hat{p}_\mu$ and $\hat{M} = \hat{j}_\mu$, of course.

Due to our choice of Lindblad operators (and $r^+ = 0$), the
procedure drastically simplifies, because exclusively jumps into
the ground state $| 0 \rangle$ or into the state $\hat{L} \, |0
\rangle$ can occur, cf.~Eqs.~(\ref{operators1}) and
(\ref{operators2}). Furthermore, since $| 0 \rangle$ is an
eigenstate of $\hat{H}_\text{eff}$, its deterministic evolutions
become trivial and feature probabilities $p_\mu(t) = 0$ and
currents $j_\mu(t) = 0$. Consequently, to the above time average
only those parts of $| \varphi(t) \rangle$ contribute, where the
deterministic propagations of $\hat{L} \, |0 \rangle$ take place,
i.e., the whole problem reduces to the evaluation of $| \psi(\tau)
\rangle = \exp(-\imath \, \hat{H}_\text{eff} \, \tau) \, \hat{L}
\, | 0 \rangle$ solely. Note that propagations with the length of
time $\tau$ appear in $| \varphi(t) \rangle$ with a frequency of
occurrence which is given by $W(\tau) = - \text{d} / \text{d}\tau
\, \langle \psi(\tau) \, | \, \psi(\tau) \rangle$,
cf.~Ref.~\cite{michel2008}. It is therefore possible to derive an
exact analytical formula for the infinite time average, namely,
\begin{equation}
\text{Tr} \{ \rho \, \hat{M} \} = \frac{1}{C} \int
\limits_0^\infty \text{d}\tau \; W(\tau) \int \limits_0^{\tau}
\text{d} \tau_1 \; \frac{\langle \psi(\tau_1) \, | \, \hat{M} \, |
\, \psi(\tau_1) \rangle}{\langle \psi(\tau_1) \, | \, \psi(\tau_1)
\rangle} \label{formula}
\end{equation}
with the constant $C = 1/ \, l^+ + \int_0^\infty \text{d}\tau \,
W(\tau) \, \tau$. Note that the dependence on the bath rate $l^+$
appears as the scaling factor $C$ solely. Thus, the remaining
crucial parameters are $l^-$ and $r^-$. For simplicity, however,
we consider the special case of a negligibly small $l^-$ (within
$\hat{H}_\text{eff}$), i.e., the case of a sufficiently weak
coupling of left bath and system. (In that case also $l^+$ becomes
small but only affects the above constant $C$. No restriction is made
for the ratio $l^-/l^+$, e.g., the temperature of the left bath
is still finite.) Only a single bath parameter remains: $r^- \equiv r$.

\begin{figure}[htb]
\centering
\includegraphics[width=0.95\linewidth]{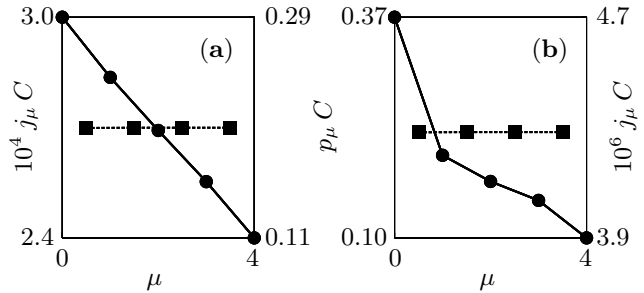}
\caption{Local probabilities $p_\mu$ (circles) and local currents
$j_\mu$ (squares) for ({\bf a}) $\lambda = 0.001$, $r = 0.0025$
and ({\bf b}) $\lambda = 8 \cdot 10^{-5}$, $r = 4 \cdot 10^{-5}$.
Remaining model parameters: $N = 5$, $n = 500$, and $\delta
\epsilon = 0.5$ ($D = 6280 \, \lambda^2$). The resulting
conductivities are ({\bf a}) $\kappa / D \approx 0.98$ and ({\bf
b}) $\kappa / D \approx 4$.} \label{N5_profiles}
\end{figure}

\begin{figure}[htb]
\centering
\includegraphics[width=0.9\linewidth]{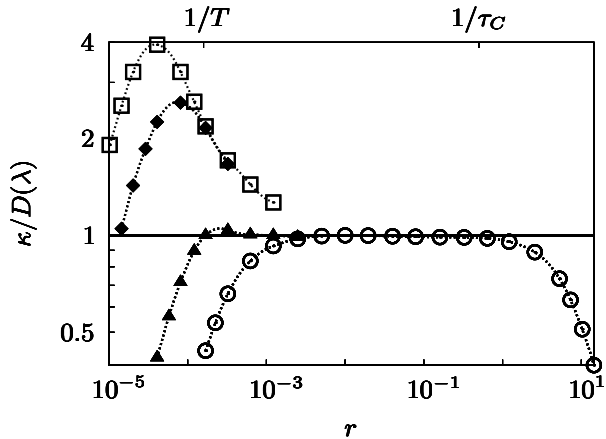}
\caption{Conductivity $\kappa$ as a function of the bath rate $r$
for $\lambda = 0.001$ (circles), $\lambda = 16 \cdot 10^{-5}$
(triangles), $\lambda = 10 \cdot 10^{-5}$ (rhombuses), and
$\lambda = 8 \cdot 10^{-5}$ (squares). Remaining model parameters:
$N = 5$, $n = 500$, and $\delta \epsilon = 0.5$ ($D = 6280 \,
\lambda^2$).} \label{N5_series_r}
\end{figure}

In practice the above integral (\ref{formula}) is approximated by
a sum over discrete time steps $\Delta \tau_i$ and with a finite
upper limit $\tau_\text{max}$. The use of not too large, i.e.,
numerically still accessible $\tau_\text{max}$ is possible,
because the function $W(\tau)$ usually decreases rapidly with
$\tau$, e.g., it decays exponentially fast. In our numerical
simulations we use $\Delta \tau_i \, D \leq 0.01$ and
$\tau_\text{max} \, D \geq 1000$. This choice typically is minimum
required for a good convergence in terms of $j_\mu$ which are
independent from $\mu$, i.e., with a deviation from their average
in the order of less than $1 \%$, at least for the following
Figs.~\ref{N5_profiles} and \ref{N5_series_r}.

Fig.~\ref{N5_profiles}a shows first numerical results for a system
with $N = 5$, where the remaining model parameters are set to well
satisfy the both conditions (\ref{conditions}) for purely
diffusive transport in the closed scenario. In the open scenario
at hand, as expected for normal behavior, we indeed observe a
strictly linear excitation profile $p_\mu$ over the whole chain
without significant effects at the boundaries $\mu = 0$ and $N-1$.
Moreover, the conductivity $\kappa = j_\mu / (p_{\mu} -
p_{\mu+1})$ turns out to agree almost perfectly with the
theoretical diffusion coefficient $D$ from
Ref.~\cite{steinigeweg2007-1}, $\kappa / D \approx 0.98$. This is
one of the main results of this work. It already indicates that
$\kappa$ may be viewed as a bulk property and not as some artifact
of the bath coupling. So far, $\kappa$ may be extracted at any
position $\mu$. But later on we will extract $\kappa$ always from
the middle of the chain.

However, a detailed investigation of $\kappa(r)$ for the same set
of model parameters is presented by the circles in
Fig.~\ref{N5_series_r}. Remarkably, $\kappa$ is found to be
independent from $r$ over almost three orders of magnitude. The
decrease of $\kappa$ in the large $r$-limit may be understood as
an effect which is due to the onset of the breakdown of the weak
coupling approximation that underlies the Lindblad form in
general, since it starts at those rates which are close to the
inverse correlation time $1/\tau_C \sim \delta \epsilon$ of the
closed system. (Note that the underlying excitation profile
$p_\mu$ is still linear except for an abrupt decline to zero at
$\mu = N-1$.)

The decrease of $\kappa$ in the small $r$-limit appears to be
contra-intuitive but nevertheless is numerically observed here.
This decrease begins at those rates which are close to the inverse
relaxation time $1/\tau_R \sim D \propto \lambda^2$ of the closed
system. Therefore one may carefully conjecture that the onset of
the decline is shifted to even smaller $r$, if only $1/\tau_R$
becomes smaller, e.g., if $\lambda$ is decreased. This conjecture
is confirmed by the triangles in Fig.~\ref{N5_series_r}.

But the further reduction of $\lambda$ eventually leads to the
violation of the second condition (\ref{conditions}), i.e., in the
closed system diffusive transport is finally expected to break
down towards ballistic behavior. As displayed in
Fig.~\ref{N5_profiles}b, such a behavior is found for the open
system at hand, too. The probability profile $p_\mu$ is highly
non-linear and the conductivity $\kappa$, now extracted in the
middle of the chain, is also much larger than the theoretical
diffusion coefficient $D$, $\kappa / D \approx 4$. (Note that the
integration of (\ref{formula}) becomes very costly, since the
function $W(\tau)$ exhibits a slowly decaying long-time tail.)

Contrary to the above diffusive case, $\kappa$ turns out to be
extremely $r$-dependent in the ballistic case, see the squares in
Fig.~\ref{N5_series_r}. (Note that these squares are evaluated for
the maximum $r$-interval which is still accessible to our
numerics). $\kappa(r)$ features a distinct peak below the inverse
correlation period $1/T = \delta \epsilon /(2 \pi \, n)$ of the
closed system. The decrease of $\kappa$ on the l.h.s.~of the peak
may be understood as the decline to zero which has already been
observed for $r \ll 1/\tau_R$. On the r.h.s.~of the peak the
decrease of $\kappa$ reaches rather fast a value which is close to
$D$ and eventually seems to reach this value but to stagnate at
it. Surprisingly, the ballistic nature of transport does not have
a significant impact on $\kappa$ here.

This observation, and all other findings in
Fig.~\ref{N5_series_r}, allow for the following interpretation:
Assume that the external bath resolves the internal dynamics on a
corresponding time scale $t = 1/r$. Then a deviation from the
transport behavior of the isolated system is to be expected, if
the bath-induced dynamics is either faster than internal
correlation times ($r > 1/\tau_C$) or slower than internal
relaxation times ($r < 1/\tau_R$). (This deviation results
in a decrease of $\kappa$.) If the bath-induced dynamics
proceeds in the possibly huge regime in between those time scales,
diffusive behavior becomes visible. However, a ballistic closed
system typically features a time scale on which the non-decaying
character of the correlation functions becomes crucial but
relaxation has not been completed yet. As already explained, for
our model such a time scale exists, if $\tau_R < T$ (i.e., if the
second condition (\ref{conditions}) is violated), precisely in
between those two times. Now, if the bath-induced dynamics is
tuned by $r$ to the latter time scale, signatures of ballistic
transport, i.e., a sharply increasing $\kappa$ appear. The larger
the latter time regime becomes, the more pronounced is the maximum
of $\kappa$ and its height may probably rise without bound
according to the degree to which the second condition
(\ref{conditions}) is violated, see Fig.~\ref{N5_series_r}. This
parameter regime is, however, numerically very costly and can with
our resources only be explored to the extend displayed in
Fig.~\ref{N5_series_r}. It is numerically even more costly to
drive the violation of the second condition (\ref{conditions}) by
increasing the length $N$ rather than decreasing $\lambda$, which
prevents us from exploring this regime directly. Nevertheless,
based on the above interpretation, we expect that the length scale
dependent transition to ballistic behavior discussed in
Ref.~\cite{steinigeweg2007-1} should become visible in the present
bath scenario, too.

To conclude, we investigated a transport scenario based on bath
coupling for a model with quantitatively well-known transport
properties. We found that a bath scenario may produce the correct
bulk conductivity but this requires a careful implementation of
the bath contacts.

\acknowledgments
We sincerely thank H.~Wichterich and M.~Michel for fruitful
discussions. Financial support by the ``Deutsche
Forschungsgemeinschaft'' is gratefully acknowledged.


\end{document}